\documentclass[conference]{IEEEtran}
\IEEEoverridecommandlockouts
\usepackage{cite}
\usepackage{amsmath,amssymb,amsfonts}
\usepackage{algorithmic}
\usepackage{graphicx}
\usepackage{textcomp}
\usepackage{xcolor}
\newtheorem{defn}{Definition}
\usepackage{bm}
\def\BibTeX{{\rm B\kern-.05em{\sc i\kern-.025em b}\kern-.08em
    T\kern-.1667em\lower.7ex\hbox{E}\kern-.125emX}}

\begin{document}

\title{A Continuous Variable Quantum Switch\\
\thanks{IT thanks NSF grant 1842559 and ORNL grant 4000178321 and KPS thanks NSF grant 2204985.}
}

\author{\IEEEauthorblockN{ Ian Tillman}
\IEEEauthorblockA{\textit{College of Optical Sciences} \\
\textit{University of Arizona}\\
Tucson, USA \\
ijtillman@arizona.edu}
\and
\IEEEauthorblockN{Thirupathaiah Vasantam}
\IEEEauthorblockA{\textit{Department of Computer Science} \\
\textit{Durham University}\\
Durham, UK \\
thirupathaiah.vasantam@durham.ac.uk}
\and
\IEEEauthorblockN{Kaushik P. Seshadreesan}
\IEEEauthorblockA{\textit{Department of Informatics \& Networked Systems} \\
\textit{School of Computing \& Information} \\
\textit{University of Pittsburgh}, Pittsburgh, USA \\
kausesh@pitt.edu}
}

\maketitle

\begin{abstract}
The continuous quadratures of a single mode of the light field present a promising avenue to encode quantum information. 
By virtue of the infinite dimensionality of the associated Hilbert space, quantum states of these continuous variables (CV) can enable higher communication rates compared to single photon-based qubit encodings. 
Quantum repeater protocols that are essential to extend the range of quantum communications at enhanced rates over direct transmission have also been recently proposed for CV quantum encodings. 
Here we present a quantum repeating switch for CV quantum encodings that caters to multiple communication flows. 
The architecture of the switch is based on quantum light sources, detectors, memories, and switching fabric, and the routing protocol is based on a Max-Weight scheduling policy that is throughput optimal. 
We present numerical results on an achievable bipartite entanglement request rate region for multiple CV entanglement flows that can be stably supported through the switch. 
We elucidate our results with the help of exemplary 3-flow networks.
\end{abstract}

\begin{IEEEkeywords}
quantum continuous variables, quantum repeater, quantum switch, maximum weight scheduling, entanglement distribution
\end{IEEEkeywords}

\section{Introduction}
The world of ``classical" computing over the past many decades has been on a fascinating journey from individual mainframes and PCs to the internet age of globally networked computers, mobile phones, IoT devices and beyond. 
Interconnecting computing devices proved to be a game-changer leading to the explosion of information technologies and applications that encompass our lives today, many of which at the outset were hardly thought to be possible, e.g., internet banking and social media.
In a similar vein, the quantum 2.0 technologies~\cite{dowlingMilburnQuantumRevolution} such as quantum computers, quantum sensors, quantum simulators, and other devices that are currently being developed promise to give rise to novel applications and to outperform classical technologies in certain types of tasks in the future. 
Arriving at this future largely hinges on interconnecting such quantum 2.0 devices to form quantum networks~\cite{WEH18}. 

The ultimate goal of quantum networking is to enable reliable exchange of quantum information between quantum 2.0 devices at high rates both locally between quantum device modules at quantum data centers as well as at global-scale between distant quantum nodes.
This includes faithful transmission of quantum bits (qubits) and distribution of quantum entanglement, which is the special type of non-classical correlations that can exist between qubits and is believed to be the workhorse behind a large number of the quantum 2.0 technologies. 
However, quantum networking is not possible over the ``classical" internet as it does not support quantum information exchange. 
It requires new networking infrastructure consisting of novel transceivers that can generate and detect quantum states, quantum memories that can store quantum information, and frequency converters that can transduce photonic quantum information into the telecom band for efficient transmission over optical fibers. 
Additionally, quantum networking requires novel quantum repeaters~\cite{MZTN15} that are not just optical amplifiers but special-purpose quantum processors to periodically re-generate quantum information in long range communications and support meaningful end-to-end rates.
Furthermore, fast and high efficiency switches are required in conjunction with repeaters to route multiple communication flows through arbitrary network topology.

Quantum information can be broadly classified into discrete and continuous variable (DV and CV) encodings.
DV quantum information refers to the information encoded in states of discrete degrees of freedom of quantum systems, e.g., the spin of an electron or nucleus, photon number of an optical mode, and polarization, path, time-bin, or frequency-bin of a photon. 
On the other hand, CV quantum information refers to quantum states of the continuous quadrature degrees of freedom of a spatial, spectral, or temporal mode of the electromagnetic field. Examples of CV quantum states include coherent states, thermal states, squeezed states, and bosonic cat states. 
A host of repeater architectures and protocols for quantum networking have been proposed and analyzed extensively for DV encodings since the 1990s based on spin qubits (e.g. in trapped ions~\cite{DLWGS22} and color-centers in diamond~\cite{RYGRHHWE19}) and single photon polarization qubits. 
Quantum switches capable of routing mutliple DV quantum entanglement have also been studied~\cite{VGNT21,thiru}. 
A switch typically operates in conjunction with repeater elements that help establish and store entanglement between the switch and the different nodes connected to its ports, followed by entanglement swapping across different sets of switch ports to support different entanglement flows. 
It was proved recently that a Max-Weight scheduling policy is throughput optimal for a switch with quantum memories that have a lifetime of one time step that caters to several DV entanglement flows, where throughput optimality refers to achieving stability of request queues associated with flows, for all feasible request rates\cite{thiru}.

Quantum repeaters have also been subsequently proposed for CV encodings. 
A prominent architecture for a CV repeater involves generating entangled two-mode quadrature squeezed states of light from Type-II spontaneous parametric downconversion (SPDC) sources, CV non-Gaussian quantum error correction by injecting and detecting single photons~\cite{diasRalph2018}, storing heralded link-level entanglement in quantum memories, and entanglement swapping by coherent detection~\cite{diasCVRepeater, ghalaiiInfiniteQS, furrerAndMunro} or by other non-Gaussian methods~\cite{kaushikPRR, furrerAndMunro}. 
For the repeater architecture with coherent detection, the end-to-end entanglement rates in a hub-and-spoke CV network architecture for different relative placements of the central repeater hub were characterized and placements that enhance the rate of each individual flow over its corresponding direct transmission capacity were identified~\cite{TRGS22}.
In the present work, we model the central repeater node as a quantum repeating CV switch that creates CV entangled links with the end nodes connected to its ports and implements routing based on a Max-Weight scheduling policy to cater to requests for the different entanglement flows between the end nodes by suitably performing entanglement swapping across its ports. 
We present numerical results on an achievable request rate region for the different end-to-end CV bipartite entanglement flows that can be supported by the switch stably. The achievable request rate region translates to an entanglement request rate region when end-to-end entanglement distillation based on the hashing protocol is considered\cite{devetak2005distillation}.

The manuscript is organized as follows: 
In Section~\ref{repeater_arch}, we describe the CV repeater architecture and protocol used in \cite{TRGS22}, which is also what we will consider in the present work.
Subsequently, in Section~\ref{setting}, we describe the model for a CV quantum repeating switch. 
Section~\ref{MWscheduling} describes the Max Weight schedule policy used by the switch. 
This is followed by results in Section~\ref{results}. 
We conclude with a discussion in Section~\ref{disc}.

\section{Continuous Variable Quantum Repeater Architecture \label{repeater_arch}}
Quantum continuous variables of a bosonic mode~\cite{gaussianQuantumInformation} are described by the Hermitian quadrature operators $\hat{q}=(\hat{a}+\hat{a}^\dagger)/\sqrt{2},$ 
$\hat{p}=(\hat{a}-\hat{a}^\dagger)/(\sqrt{2}i)$, where $\hat{a}^\dagger, \hat{a}$ denote the creation and annihilation operators of the mode and we have chosen $\hbar=1$. 
The quintessential entangled state in CV is the two-mode squeezed vacuum state (TMSV) $|{\chi}\rangle_{AB}$, which is expressed in the photon number basis as
$|{\chi}\rangle_{AB} = \sqrt{1-\chi^2} \sum_{n=0}^{\infty} \chi^n |{n,n}\rangle_{AB},$
where $0 \leq \chi < 1$. 
The TMSV state exhibits a reduced, or `squeezed', variance of either the difference of the position quadratures of the two modes or the sum of their momentum quadratures at the expense of an increased variance of the other. 
When one of the two modes of a TMSV state is transmitted through a lossy channel of transmissivity $\eta$, in the limit $\chi\rightarrow 1$ (which is the infinite energy limit) the distillable entanglement of the end-to-end state coincides with the entanglement distribution capacity of the channel~\cite{PLOB} given by
\begin{align}
C_{\text{direct}}(\eta)= -\log_2(1-\eta)\ \textrm{ebits/mode},\label{PLOB_ebits/mode}
\end{align}
which in the limit $\eta\ll 1$ is $C(\eta)\propto \eta$. (Here an ebit refers to a maximally entangled pair of qubits.)

\begin{figure}[htbp]
\centerline{\includegraphics[width=0.99\linewidth]{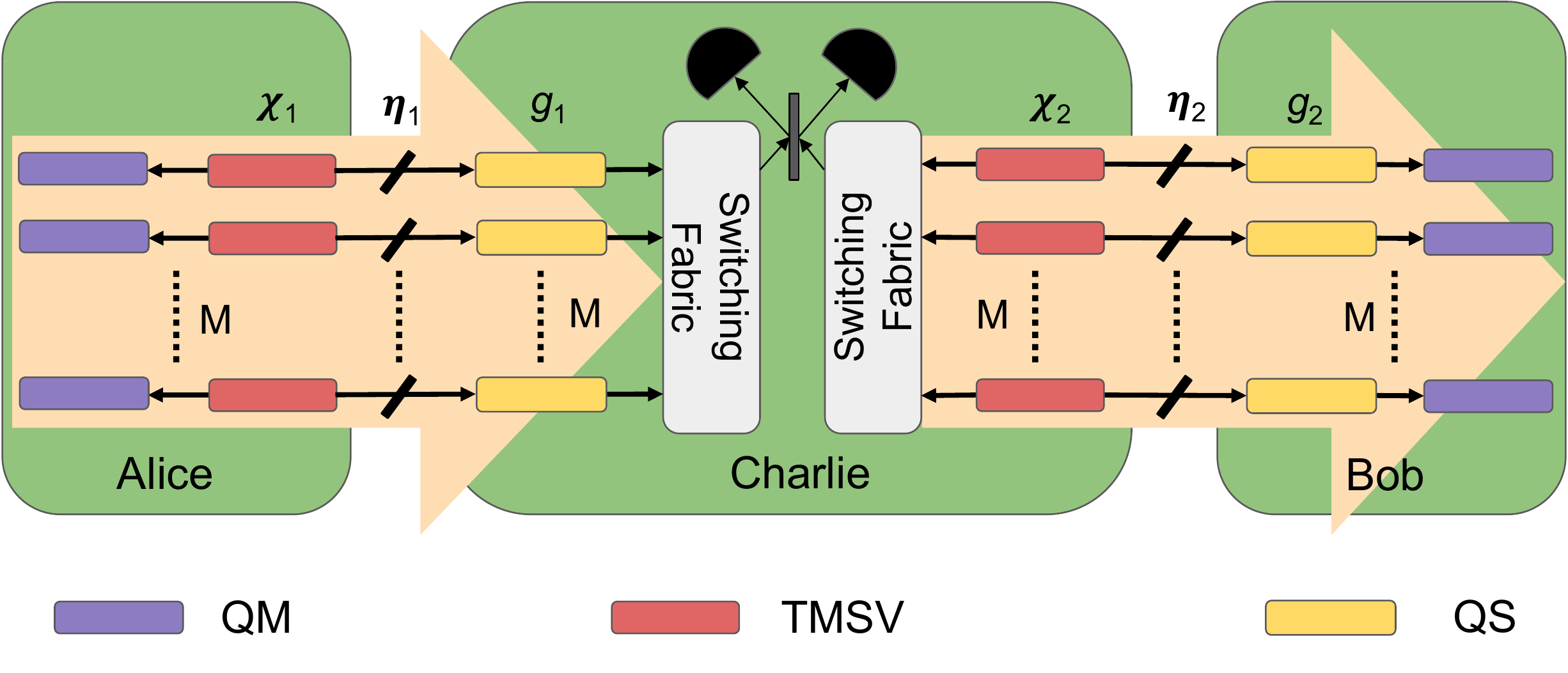}}
\caption{A CV quantum repeater protocol for entanglement distribution. The repeaters rely on multiplexed distribution of two-mode squeezed vacuum (TMSV) states across each link (i.e., lossy links between the end nodes Alice and Bob, and the repeater node Charlie), quantum scissor (QS)-based probabilistic heralded noiseless linear amplification to mitigate the effects of photon loss modeled by finite transmissivities $\eta_i, i\in\{1,2\}$ (on either side of Charlie), and dual homodyne detection based entanglement swapping at Charlie to connect the heralded and successfully error corrected entangled links on either side of Charlie. The end-to-end entanglement is then stored in quantum memories (QM). Notice the Orange arrows (background) that are aligned in the same direction (left to right) indicating the relative orientation of the links (TMSV tail and QS head) being connected at Charlie (head to tail).}
\label{fig1}
\end{figure}

A recently proposed and widely studied repeater architecture~\cite{diasCVRepeater} for CV that can help surpass the direct transmission capacity of (\ref{PLOB_ebits/mode}) for entanglement distribution between end nodes is shown in Fig.~\ref{fig1}. 
It also involves distributing TMSV states across each individual link in a repeater network. 
The repeater employs so-called non-deterministic noiseless linear amplification (NLA)~\cite{ralphLundQSNLA} (with an associated gain parameter $g$) at the repeaters to perform non-Gaussian CV quantum error correction on the lossy TMSV states across individual links~\cite{NFC09}. 
The concept of NLA comes under the class of `immaculate' quantum amplifiers~\cite{PJCC13} that can amplify CV states without adding any noise to the system, but only succeed in doing so probabilistically. 
In order to boost the success probability of entanglement generation in each time step, the repeaters multiplex their entanglement generation attempts ($M$ fold multiplexing) across each link. 
In the limit of $\eta\ll1$, the NLA can be approximately realized using an optical state engineering primitive known as the `quantum scissor'~\cite{peggQuantumScissor}, which involves single photon injection and detection. 
The repeater relies on coherent dual homodyne detection~\cite{covarianceMatrixAlgebra}, which is a Gaussian measurement, for entanglement swapping at the repeaters to connect adjacent links and have long range end-to-end entanglement, which can then be stored in the quantum memories at the end nodes. 

The above multiplexed CV repeater architecture based on TMSV and quantum scissor NLA in the limit $\eta\ll 1$ can support an end-to-end entanglement rate that scales as $\propto \eta^{3/4}$~\cite{TRGS22}, thus surpassing the direct transmission capacity of (\ref{PLOB_ebits/mode}). 
It is important to note that the said rate-loss scaling is achieved when the NLA-end of one link is connected to the TMSV-end of the other, as indicated by the aligned orange shaded arrows in Fig.~\ref{fig1}, while the other relative orientations such as connecting 2 NLA ends or 2 TMSV ends result in strictly worse end-to-end entanglement rates. 
For the different orientations of two adjacent repeater links and for different transmissivities $\eta_i,\ i\in\{1,2\}$ in the two links, the end-to-end entanglement rates were determined in~\cite{TRGS22} by evaluating an information theoretic lower bound on the asymptotic distillable entanglement of the end-to-end state, known as the reverse coherent information. 
Using the entanglement rates as a function of the different transmissivities~\cite{TRGS22} further gave the end-to-end entanglement rates in a hub-and-spoke quantum network with a quantum repeater as the hub node for different placements of the hub relative to the end nodes, thus identifying placements that help surpass the direct transmission capacity of (\ref{PLOB_ebits/mode}) for each individual entanglement flow through the repeater hub.



\section{Continuous Variable Quantum Switch Model\label{setting}}
\begin{figure}[htbp]
\centerline{\includegraphics[width=0.99\linewidth]{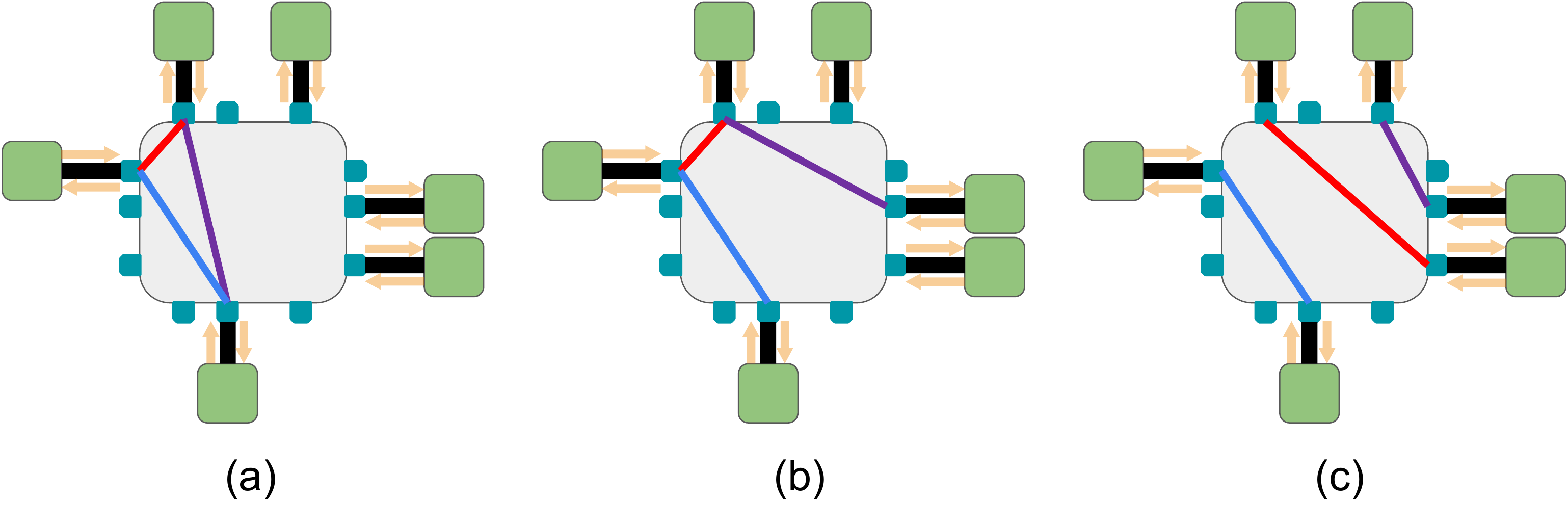}}
\caption{A CV quantum repeating switch architecture. Each end node that is connected to the switch attempts heralded CV entanglement generation with the switch at each time step using multiplexed two-mode squeezed vacuum (TMSV) sources and quantum scissors (QS) depicted in Fig.~\ref{fig1}. The attempts happen in both orientations, i.e., TMSV at the end node and QS at the switch and vice versa, as indicated by the pair of Orange arrows. (a), (b), (c) describe the 3 possible scenarios associated with 3 bipartite entanglement flows (denoted by Red, Blue and Purple lines) at the switch. We expand on these possibilities in sec. \ref{results}}
\label{fig2}
\end{figure}

Here, we describe our model for a quantum repeating switch that can route multiple CV entanglement flows. 
For simplicity, we restrict ourselves to bipartite entanglement flows. 
The model is based on a hub-and-spoke network topology with the switch as the hub node and the end nodes as the spokes as described in Fig.~\ref{fig2}. 
In every time step, the central repeater hub node attempts CV entanglement generation with each end node along the spokes, or `links', using TMSV states and QS NLA-based error correction over $M$ multiplexed channels. 
Notably, the entanglement generation across each link is attempted along both link orientations, i.e., half the multiplexed channels use TMSV sources at the end node and the QSs at the switch port, while the other half use TMSV sources at the switch and the QSs at the end node. 
The rationale behind this is to have a succcessfully generated entanglement across a link be equally likely to be either possible orientation. This allows the switch to connect aligned links capable of providing larger entanglement rates when possible, and do so impartially between the different entanglement flows through the switch
By suitably tuning the gain parameter of the QS-based NLA the probability of heralding a successfully error corrected CV entangled state across a link, $P_\mathrm{NLA}$, can be made to scale as $\eta^{1/4}$ in the limit of $ \eta \ll 1 $. 
By attempting entanglement generation across $M$ multiplexed channels, the probability of successfully heralding at least 1 error corrected CV entanglement state across a link is given by $p=1-(1-P_\mathrm{NLA})^M$.
The switch then picks a request from the queue that can be serviced with the links that were able to generate entanglement in that particular time step and performs a dual homodyne detection based entanglement swap to entangle the two end nodes.
The choice of which ports to connect via entanglement swapping in any given time step is driven by a throughput optimal routing policy described in Section~\ref{MWscheduling}. 
A point to note about the dual homodyne detection based CV entanglement swap operation is that it is deterministic unless we choose to perform post-selection to optimize the end-to-end entangled state, which for simplicity we do not do here.
In other words, the entanglement swap success probability of the CV quantum switch is taken to be $q=1$.

Note that our model assumes that both the end nodes and the switch have unconstrained quantum resources (quantum sources, detectors and memories) and share a synchronized clock. 
We assume that the lifetime of the quantum memories at the switch (surpressed in Fig. \ref{fig1}) equals the duration of a time step of operating the switch, whereas the lifetime of the memories at the end users are assumed to be several time steps long, enabling the hashing protocol.
This simplifies our model and multiple entangled states heralded during one step do not get carried over to subsequent time steps.


\section{Throughput Optimal Routing \label{MWscheduling}}
We assume that requests of each entanglement flow type, arrive at the switch according to an i.i.d. process with $\lambda_i$ denoting the average number of type $i$ requests arriving in each time step. A type $i$ request demands the switch to create an entanglement shared by a set of users, $U_i$. We make an assumption that requests of each type are processed in the order they arrive at the switch and there is a buffer for each type of requests with infinite classical memory to store unserved requests. Let $Q_i(n)$ be the number of requests of type $i$ waiting in the queue corresponding to type $i$ requests at the beginning of time step $n$. We use $T_j(n)$ to indicate whether an elementary entanglement is available or not on link $j$ in time step $n$; if an elementary entanglement is present then $T_j(n)=1$ and if no elementary entanglement is present then $T_j(n)=0$.

Since we pick at most one elementary entanglement per time step along each link, if there are requests of multiple types waiting for service then there exists a competition among requests to use available elementary entanglements. The switch should make scheduling decisions to decide which requests to be processed using available elementary entanglements in such a way that requests are processed as quickly as possible. As in \cite{thiru}, the switch is considered to process requests in each time step based on a selected matching $\mathbf{r}=[r_1,r_2,...,r_k]$ which indicates that the switch attempts to serve $r_i$ requests of type $i$ by performing entanglement swapping operations. The term matching is defined as follows:
\begin{defn}{\textit{Matching:}}
A vector $\mathbf{r}=[r_1,r_2,...,r_k]$ with $r_i\in\{0,1\}$ for $1\leq i\leq K$ is called matching if for every user $j$, we have
\begin{equation}
    \sum_{m=1,j\in U_m}^Kr_m\leq 1.
\end{equation}
\end{defn}
The above equation implies that the number of requested elementary entanglements as per the matching $\mathbf{r}$, that are shared between user $j$ and the switch should be at most one.

We use the Max-Weight scheduling policy of \cite{thiru} to make scheduling decisions at the switch that decides which matching should be selected from the set of all feasible matchings. We formally define the scheduling policy below:
\begin{defn}{\textit{Max-Weight scheduling policy}:}
According to the max-weight policy, a matching $\mathbf{W}(n)=[W_1(n),\cdots,W_K(n)]$ is selected in time step $n$ if
\begin{equation}
\label{eq:maxweight}
    \mathbf{W}(n)=\arg\max_{\mathbf{r}\in\mathcal{M}}\sum_i r_iq_i Q_i(n) I_{\{T_j(n)>0,\forall j\in U_i\}},
\end{equation}
where $\mathcal{M}$ denotes the set of all matchings and $q_i$ is the success probability of entanglement swapping operations of type $i$ requests. 
\end{defn}
In \cite{thiru}, the capacity region of a switch was found, defined as the set of request rates $\bm{\lambda}=[\lambda_1,\cdots,\lambda_K]$ for which there can exist a scheduling policy that stabilizes the switch, guaranteeing finite average waiting times for requests. If $\bm{\lambda}$ is not in the capacity region, then no scheduling policy can stabilize the switch. It was also proved in \cite{thiru} that the Max-Weight policy defined in \eqref{eq:maxweight} is throughput optimal, that is, it stabilizes the switch for all request arrival rates that belong to the capacity region. If the switch is stable under a scheduling policy for a rate vector $\bm{\lambda}$, then in the stationary regime, the departure rate of requests leaving the switch after getting service coincides with $\bm{\lambda}$. One challenge is that, it is difficult to verify whether a given rate vector $\mathbf{\lambda}$ satisfies the necessary conditions of \cite{thiru} or not to judge stability of the switch. In the next section, for a switch catering to three entanglement flows, we present results that are useful to conclude whether a switch will be stable or not for given $\mathbf{\lambda}$.

\section{Results \label{results}}

Here we present our results on the bipartite entanglement rate region for multiple CV entanglement flows through the CV quantum switch that can be stably supported. 
The results are shown in terms of the rate region for the requests corresponding to the different entanglement flows types that can be stably met by the Max-Weight scheduling policy based routing. 
An achievable end-to-end entanglement rate follows from an achievable request rate given as the latter scaled by the reverse coherent information of the end-to-end entanglement state, which is a lower bound on the distillable entanglement of the state in the asymptotic limit of many copies of the state \cite{devetak2005distillation}. 

We present our results for the example of 3-flow networks with numerical plots of the request rate regions that are stably supported by the Max-Weight scheduling policy-based routing. 
Consider the three distinct scenarios that are possible in 3-flow networks as described in Fig.~\ref{fig2}, namely (a) where all 3 flows contend with each other in each time step, (b) where 2 out of the 3 flows (Blue and Purple) are disjoint and do not contend with each other, and (c) where all 3 flows are disjoint and thus there is no contention. 
The request rate regions for each of these scenarios are functions of the entanglement swapping success probabilities $q_j$, where $j\in\{ 1,2,3 \}$, and the success probabilities of heralding entanglement along each link, $p_i$, where $i\in\{1,2,3\}$ for case (a), $i\in\{1,2,3,4\}$ for case (b), and $i\in\{1,2,3,4,5,6\}$ in case (c). 
The probabilities $p_i$ are functions of the link transmissivities $\eta_i$, the squeezing parameters $\chi_i$, and the multiplexing parameters $M_i$.
For the sake of simplicity in depicting the results we will use $M_i = 1/P_{\mathrm{NLA},i}$, which in the limit $P_{\mathrm{NLA},i}\rightarrow 0$ (i.e. when $\eta_i\rightarrow 0$) we have $p_i=1-(1-P_{\mathrm{NLA},i})^{M_i} = 1-(1-P_{\mathrm{NLA},i})^{(1/P_{\mathrm{NLA},i})} \rightarrow 1 - 1/e \approx 0.632$, approaching from above.
This gives a universal and scale-independent minimum value for the link generation success probabilities $p_i$ that also gives a near-optimal entanglement bitrate per mode \cite{TRGS22}. Along with our assumption of entanglement swapping probabilities $q_j = 1$ for all flows $j\in\{1,2,3\}$, this allows for a greatly simplified discussion.

For the scenario in Fig.~\ref{fig2} (a) where all the 3 flows contend with each other, ignoring the orientation of the successfully generated entanglements along each link, a request rate region that can be stably supported by the Max-Weight scheduling policy based routing at the switch is given by:
\begin{align}
\lambda_1+\lambda_2+\lambda_3 &\leq p^3 + 3(1-p)p^2\label{eqn:rateregion1_1}\\
\lambda_i+\lambda_j &\leq p^3 + 2(1-p)p^2, \ i,j\in\{1,2,3\}, \ i\neq j\label{eqn:rateregion1_2}\\
\lambda_i & \leq p^3 + (1-p)p^2 = p^2, \ i\in\{1,2,3\},\label{eqn:rateregion1_3}
\end{align}

\noindent where $\lambda_1,\ \lambda_2,\ \lambda_3$ are the request rates corresponding to the 3 contending flows. 
These bounds can be understood from a combinatorial perspective, as in order for the switch to be stable it must be able to service requests at least as fast as they come in. When considering a set of $1 \leq N \leq 3$ flows the sum of their request rates must be less than the probability that at least one of them can be serviced, which occurs when all elementary links are generated, which has probability $p^3$, or exactly two elementary links are formed, which has probability $N(1-p)p^2$.
A stable request rate region for the scenario in Fig.~\ref{fig2} (b) is similar to the above with the exception of the constraints on the sum of the two disjoint flows (Blue and Purple) $\lambda_1$ and $\lambda_3$ now being omitted.
Finally, the rate region for the scenario in Fig.~\ref{fig2} (c) is simply given by $\lambda_i \leq p^2, \ i\in\{1,2,3\}$. 

\begin{figure}[htbp]
\centerline{\includegraphics[width=1\linewidth]{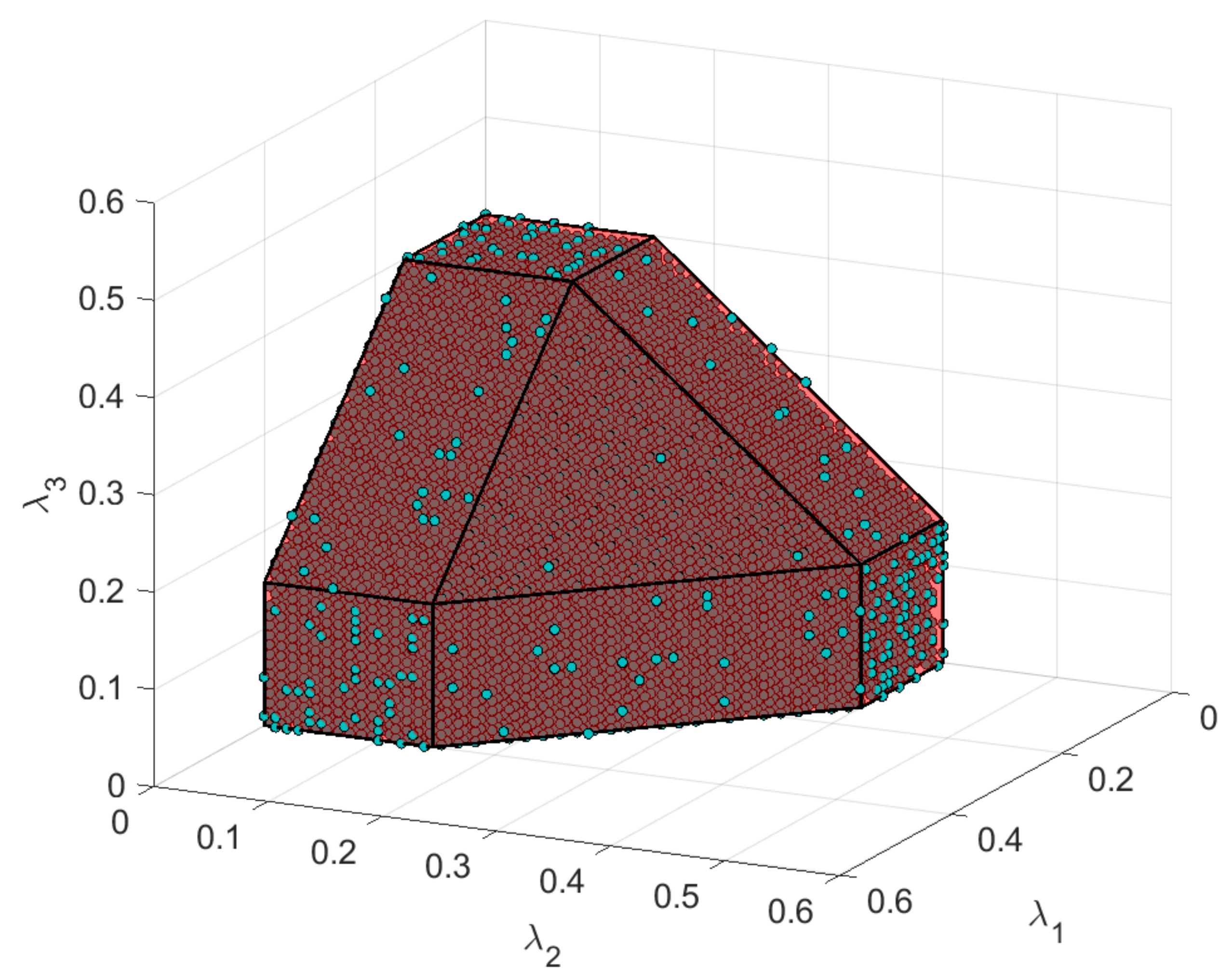}}
\caption{Request rate region for 3 overlapping flows as described in Fig.~\ref{fig2} (a) that can be stably supported by the CV repeating switch with link generation probabilities $p=0.632$ and deterministic entanglement swapping.}
\label{fig3}
\end{figure}

\begin{figure}[htbp]
\centerline{\includegraphics[width=1\linewidth]{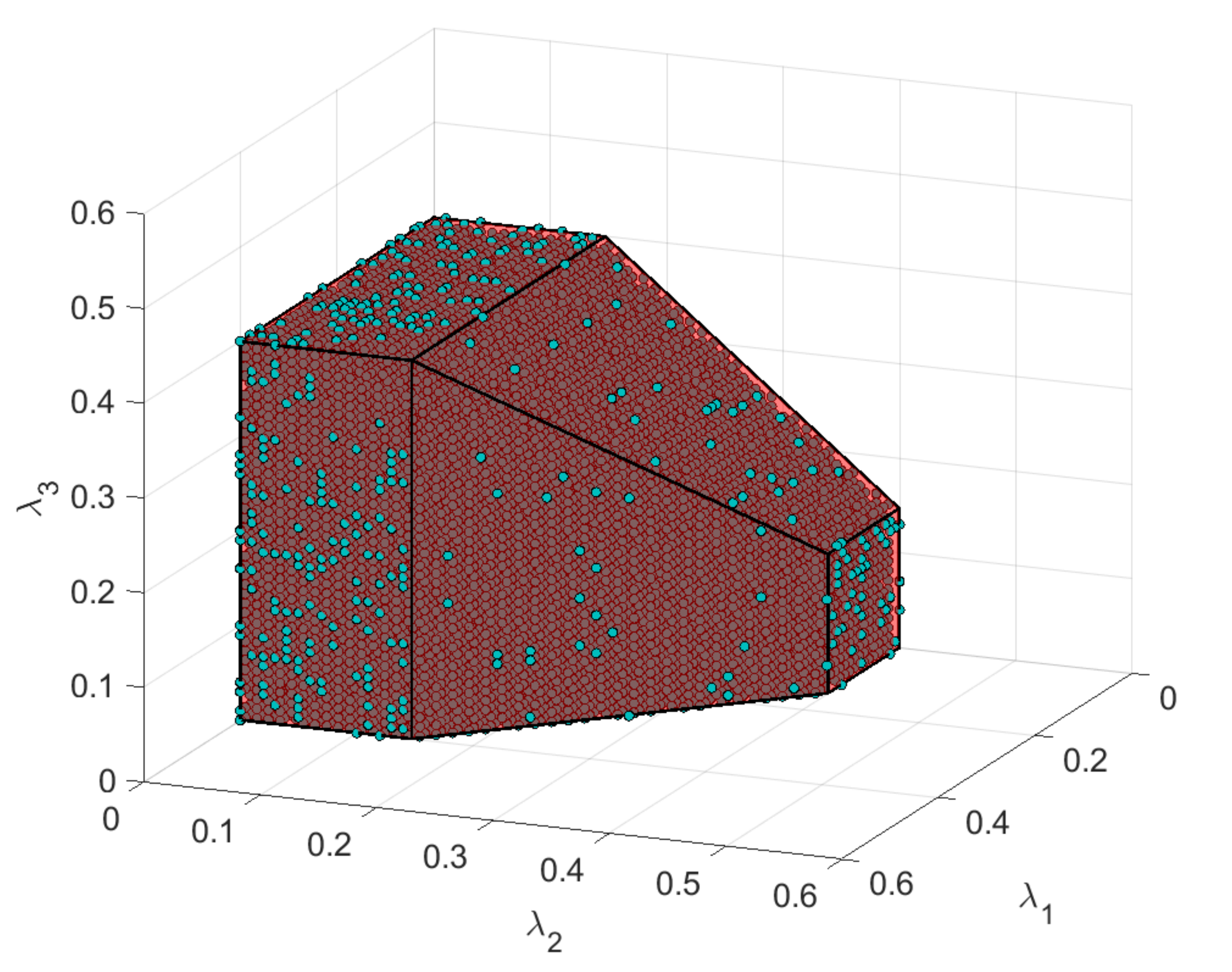}}
\caption{Request rate region for the 3-flow network scenario described in Fig.~\ref{fig2} (b) where 2 out of the 3 flows are disjoint (Blue and Purple flows, with the associated request rates being $\lambda_1$ and $\lambda_3$, respectively), that can be stably supported by the CV repeating switch with link generation probabilities $p=0.632$ and deterministic entanglement swapping. }
\label{fig4}
\end{figure}

\begin{figure}[htbp]
\centerline{\includegraphics[width=1.06\linewidth]{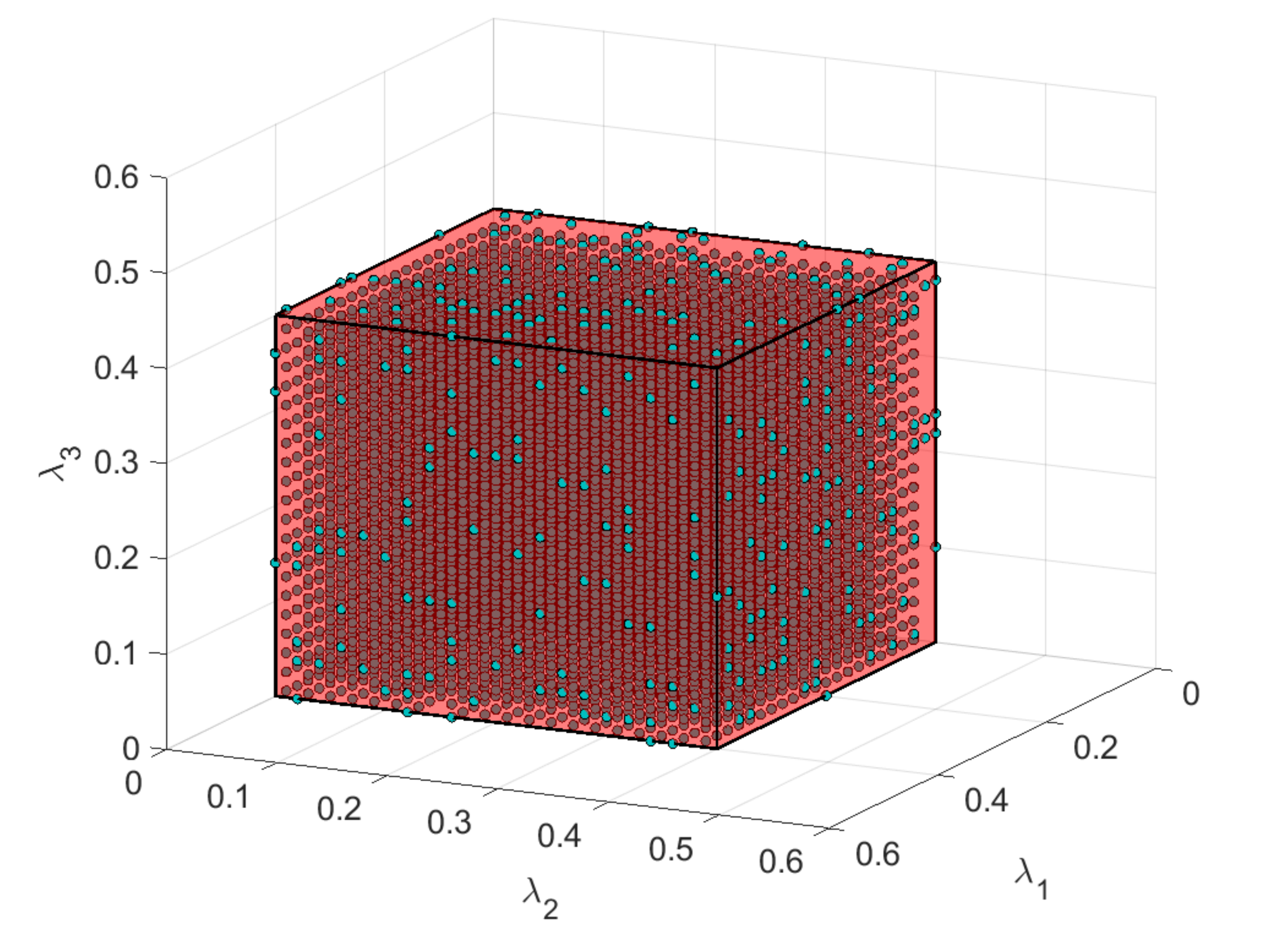}}
\caption{Request rate region for 3 disjoint flows as described in Fig.~\ref{fig2} (c) that can be stably supported by the CV repeating switch with link generation probabilities $p=0.632$ and deterministic entanglement swapping.}
\label{fig5}
\end{figure}

These rate regions for $p=0.632$ are plotted in Figs.~\ref{fig3}-\ref{fig5} along with simulated rate regions. The simulations were done for $0 \leq \lambda_i \leq 1,\  i\in\{1,2,3\}$ with step sizes of $\Delta\lambda = 0.005$ and $10^5$ time steps each.
A rate vector is considered 'stable' if a linear regression of the queue size over all time steps has a slope less than some threshold, here taken to be $10^{-4}$.
Because these simulations were done with a finite number of time steps we should expect some of the points just outside the actual rate region to be mislabeled as stable due to the average queue size diverging too slowly in that particular simulation.

\begin{figure}[htbp]
\centerline{\includegraphics[width=1.1\linewidth]{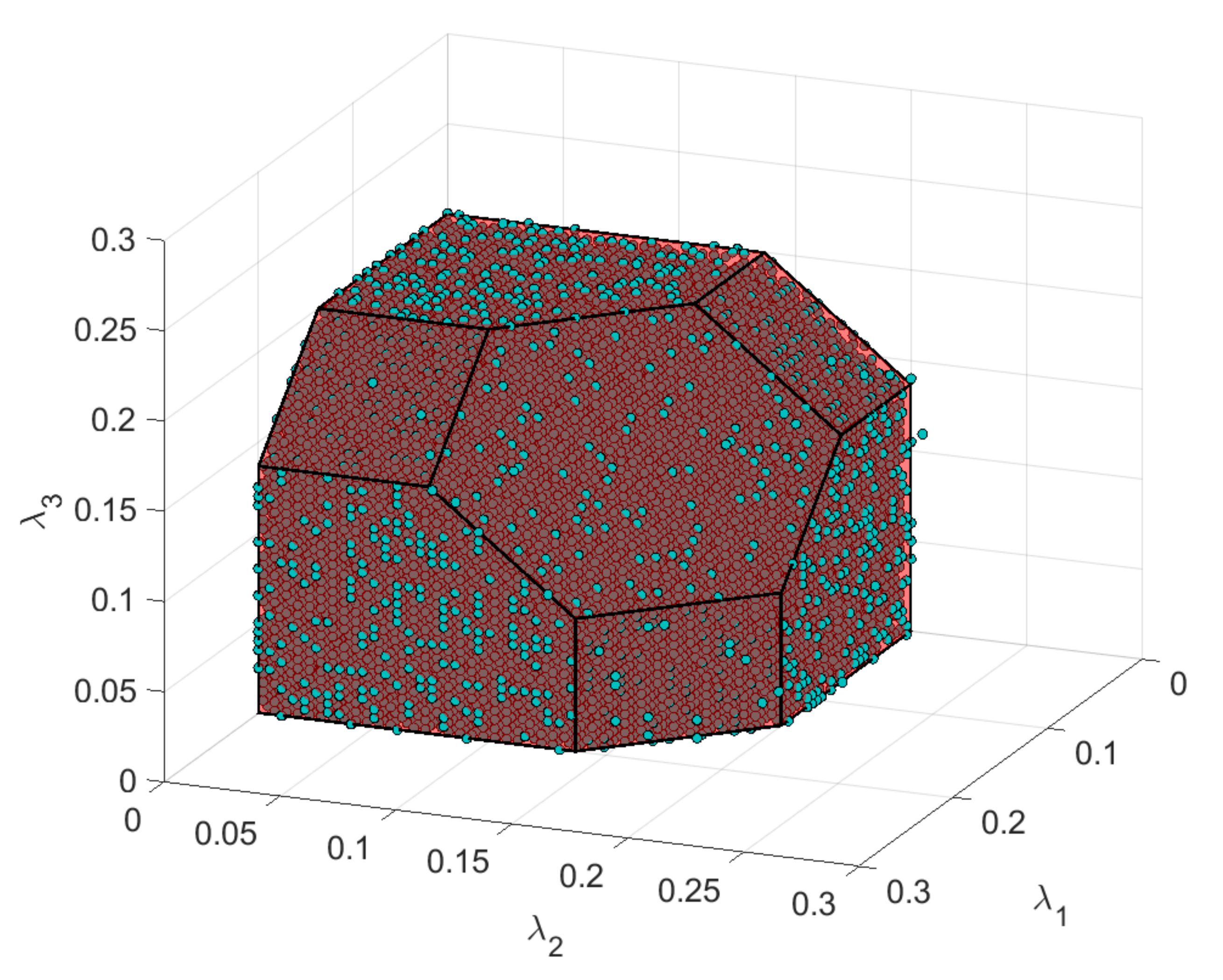}}
\caption{Request rate region for 3 overlapping flows as described in Fig.~\ref{fig2} (a) that can be stably supported by the CV repeating switch with link generation probabilities $p=0.632$ and deterministic entanglement swapping, where the relative orientations of the successfully heralded links are taken into consideration and aligned links alone are connected by the switch.}
\label{fig6}
\end{figure}

Since connecting two links that are aligned as shown in Fig.~\ref{fig1} results in an end-to-end entangled state of a higher reverse coherent information than any other orientation, we can additionally instruct the switch to take the orientations of the links into consideration in addressing routing requests. 
For example, in the case of the scenario in Fig.~\ref{fig2} (a) with contending flows and with the orientation of the successfully generated entanglements taken into account by the scheduling policy, the new request rate region is given by
\begin{align}
\lambda_1+\lambda_2+\lambda_3 &\leq 3p^2 / 2 - 3p^3 / 4\\
\lambda_i+\lambda_j &\leq p^2 - p^3 / 4, \ i,j\in\{1,2,3\}, \ i\neq j\\
\lambda_i & \leq p^2/2, \ i\in\{1,2,3\}.
\end{align}
This rate region is plotted in Fig.~\ref{fig6}. 
These are derived in the same way as equations \ref{eqn:rateregion1_1}-\ref{eqn:rateregion1_3}, except we also take into account that two links must have opposite parity to service a flow.
For example the rate of any single flow, $\lambda_i$, must not exceed the probability of the two associated elementary links succeeding, $p^2$, multiplied by $1/2$ to account for the probability that the two links have opposite parity.
When looking at the sum of two flows, $\lambda_i + \lambda_j$, $i \neq j$, this will succeed when all links succeed they are not all the same parity ($ \frac{3}{4}p^3$) or when one flow's links succeed with opposite parity and the third link fails ($ \frac{1}{2}p^2 (1-p) $), which can happen in two ways. Therefore $\lambda_i + \lambda_j \leq \frac{3}{4}p^3 + 2p^2 (1-p) \frac{1}{2} = p^2 - p^3/4$.
Finally, the sum of all three rates can be analyzed in the same way but there are now three ways in which a single flow's links succeed with opposite parity while the third link fails. So $\lambda_1 + \lambda_2 + \lambda_3 \leq \frac{3}{4}p^3 + 3p^2(1-p)\frac{1}{2} = 3p^2/2 - 3p^3 / 4$.
Though the request rates supported come down after we require opposite parity, the higher reverse coherent information of the end-to-end entangled states may result in higher entanglement rates. 
Similar request regions with the relative orientations being taking in account for the links can also be arrived at for the other two scenarios described in Fig.~\ref{fig2}.

This combinatorial approach of defining a boundary plane for each subset of the request rates can be done for any switch scenario. This includes scenarios where the link generation probabilities $p_i$ are not identical, scenarios where the entanglement fusion probabilities $q_i$ are not unity, or scenarios where the switch offers to service requests for multipartite entanglement. The plane associated with each subset of request rates are calculated by bounding the probability that at least one of those rates can be serviced, which is the same procedure as outlined in the two examples above.


\section{Discussion \label{disc}}
Our results can be improved a few obvious ways. 
First of all, for practical relevance, constrained operation of CV switches with limited resources such as TMSV sources, single photon sources and detectors, homodyne detectors and quantum memories need to be considered. 
In the resource-constrained case, it should also be interesting to consider spatial vs time multiplexed operation of the CV repeater links.
This more practical analysis should also take into account the fidelity decay inherent to the quantum memories being used, particularly when it comes to end nodes storing the final entangled states if the application in question requires entanglement distillation.
Secondly, the end-to-end entanglement rates could be further enhanced by relaxing the constraint that unused entangled states successful heralded along links be dropped after each time step and instead by modeling finite but longer lifetime for the quantum memories than the time step of the switch. 
Thirdly, it should be interesting to extend the utility of the switch to go beyond just bipartite entanglement flows and to cater to multipartite flows. 
This would require the switch to perform multimode entangled Gaussian fusion operations on the QS-NLA-enhanced entangled states along different links, which generalize the dual homodyne detection~\cite{covarianceMatrixAlgebra}. 
It should also be interesting to consider more general non-Gaussian fusion operations directly operating on the lossy TMSV state-based links, e.g., generalizing the recently proposed CV repeater protocol that uses a generalized QS to accomplish both error correction and entanglement swapping~\cite{WGHR21}.
We leave these tasks for future work. 

In summary, we presented a model for a continuous variable quantum switch where multiple end nodes share CV entanglement links based on TMSV and QS-NLA-based error corrected with the switch, and the switch routes the multiple entanglement flows between the end nodes using a Max-Weight scheduling-based routing protocol that has been shown to be throughput optimal~\cite{thiru}. 
We presented numerical results on request rate regions for 3-flow networks that can be stably supported by the CV switch. Our work paves the way to determine the capabilities of general CV quantum networks of arbitrary topology, where the switches implement a routing protocol based on a well-defined scheduling policy.




\bibliography{references}

\begin{thebibliography}{10}
\providecommand{\url}[1]{#1}
\csname url@samestyle\endcsname
\providecommand{\newblock}{\relax}
\providecommand{\bibinfo}[2]{#2}
\providecommand{\BIBentrySTDinterwordspacing}{\spaceskip=0pt\relax}
\providecommand{\BIBentryALTinterwordstretchfactor}{4}
\providecommand{\BIBentryALTinterwordspacing}{\spaceskip=\fontdimen2\font plus
\BIBentryALTinterwordstretchfactor\fontdimen3\font minus
  \fontdimen4\font\relax}
\providecommand{\BIBforeignlanguage}[2]{{%
\expandafter\ifx\csname l@#1\endcsname\relax
\typeout{** WARNING: IEEEtran.bst: No hyphenation pattern has been}%
\typeout{** loaded for the language `#1'. Using the pattern for}%
\typeout{** the default language instead.}%
\else
\language=\csname l@#1\endcsname
\fi
#2}}
\providecommand{\BIBdecl}{\relax}
\BIBdecl

\bibitem{dowlingMilburnQuantumRevolution}
J.~P. Dowling and G.~J. Milburn, ``Quantum technology: the second quantum
  revolution,'' \emph{Philosophical Transactions of the Royal Society of
  London. Series A: Mathematical, Physical and Engineering Sciences}, vol. 361,
  no. 1809, pp. 1655--1674, 2003.

\bibitem{WEH18}
\BIBentryALTinterwordspacing
S.~Wehner, D.~Elkouss, and R.~Hanson, ``Quantum internet: A vision for the road
  ahead,'' \emph{Science}, vol. 362, no. 6412, p. eaam9288, 2018. [Online].
  Available: \url{https://www.science.org/doi/abs/10.1126/science.aam9288}
\BIBentrySTDinterwordspacing

\bibitem{MZTN15}
W.~J. Munro, K.~Azuma, K.~Tamaki, and K.~Nemoto, ``Inside quantum repeaters,''
  \emph{IEEE Journal of Selected Topics in Quantum Electronics}, vol.~21,
  no.~3, pp. 78--90, 2015.

\bibitem{DLWGS22}
\BIBentryALTinterwordspacing
P.~Dhara, N.~M. Linke, E.~Waks, S.~Guha, and K.~P. Seshadreesan, ``Multiplexed
  quantum repeaters based on dual-species trapped-ion systems,'' \emph{Phys.
  Rev. A}, vol. 105, p. 022623, Feb 2022. [Online]. Available:
  \url{https://link.aps.org/doi/10.1103/PhysRevA.105.022623}
\BIBentrySTDinterwordspacing

\bibitem{RYGRHHWE19}
\BIBentryALTinterwordspacing
F.~Rozepedek, R.~Yehia, K.~Goodenough, M.~Ruf, P.~C. Humphreys, R.~Hanson,
  S.~Wehner, and D.~Elkouss, ``Near-term quantum-repeater experiments with
  nitrogen-vacancy centers: Overcoming the limitations of direct
  transmission,'' \emph{Phys. Rev. A}, vol.~99, p. 052330, May 2019. [Online].
  Available: \url{https://link.aps.org/doi/10.1103/PhysRevA.99.052330}
\BIBentrySTDinterwordspacing

\bibitem{VGNT21}
G.~Vardoyan, S.~Guha, P.~Nain, and D.~Towsley, ``On the stochastic analysis of
  a quantum entanglement distribution switch,'' \emph{IEEE Transactions on
  Quantum Engineering}, vol.~2, pp. 1--16, 2021.

\bibitem{thiru}
\BIBentryALTinterwordspacing
T.~Vasantam and D.~Towsley, ``{A throughput optimal scheduling policy for a
  quantum switch},'' in \emph{Quantum Computing, Communication, and Simulation
  II}, P.~R. Hemmer and A.~L. Migdall, Eds., vol. 12015, International Society
  for Optics and Photonics.\hskip 1em plus 0.5em minus 0.4em\relax SPIE, 2022,
  pp. 14 -- 23. [Online]. Available: \url{https://doi.org/10.1117/12.2616950}
\BIBentrySTDinterwordspacing

\bibitem{diasRalph2018}
J.~Dias and T.~C. Ralph, ``Quantum error correction of continuous-variable
  states with realistic resources,'' \emph{Physical Review A}, vol.~97, no.~3,
  p. 032335, 2018.

\bibitem{diasCVRepeater}
J.~Dias, M.~S. Winnel, N.~Hosseinidehaj, and T.~C. Ralph, ``Quantum repeater
  for continuous-variable entanglement distribution,'' \emph{Physical Review
  A}, vol. 102, no.~5, p. 052425, 2020.

\bibitem{ghalaiiInfiniteQS}
M.~Ghalaii and S.~Pirandola, ``Capacity-approaching quantum repeaters for
  quantum communications,'' \emph{Physical Review A}, vol. 102, no.~6, p.
  062412, 2020.

\bibitem{furrerAndMunro}
F.~Furrer and W.~J. Munro, ``Repeaters for continuous-variable quantum
  communication,'' \emph{Physical Review A}, vol.~98, no.~3, p. 032335, 2018.

\bibitem{kaushikPRR}
K.~P. Seshadreesan, H.~Krovi, and S.~Guha, ``Continuous-variable quantum
  repeater based on quantum scissors and mode multiplexing,'' \emph{Physical
  Review Research}, vol.~2, no.~1, p. 013310, 2020.

\bibitem{TRGS22}
\BIBentryALTinterwordspacing
I.~J. Tillman, A.~Rubenok, S.~Guha, and K.~P. Seshadreesan, ``Supporting
  multiple entanglement flows through a continuous-variable quantum repeater,''
  2022. [Online]. Available: \url{https://arxiv.org/abs/2203.07965}
\BIBentrySTDinterwordspacing

\bibitem{devetak2005distillation}
I.~Devetak and A.~Winter, ``Distillation of secret key and entanglement from
  quantum states,'' \emph{Proceedings of the Royal Society A: Mathematical,
  Physical and engineering sciences}, vol. 461, no. 2053, pp. 207--235, 2005.

\bibitem{gaussianQuantumInformation}
C.~Weedbrook, S.~Pirandola, R.~Garc{\'\i}a-Patr{\'o}n, N.~J. Cerf, T.~C. Ralph,
  J.~H. Shapiro, and S.~Lloyd, ``Gaussian quantum information,'' \emph{Reviews
  of Modern Physics}, vol.~84, no.~2, p. 621, 2012.

\bibitem{PLOB}
S.~Pirandola, R.~Laurenza, C.~Ottaviani, and L.~Banchi, ``Fundamental limits of
  repeaterless quantum communications,'' \emph{Nature communications}, vol.~8,
  no.~1, pp. 1--15, 2017.

\bibitem{ralphLundQSNLA}
T.~C. Ralph and A.~Lund, ``Nondeterministic noiseless linear amplification of
  quantum systems,'' in \emph{AIP Conference Proceedings}, vol. 1110,
  no.~1.\hskip 1em plus 0.5em minus 0.4em\relax American Institute of Physics,
  2009, pp. 155--160.

\bibitem{NFC09}
\BIBentryALTinterwordspacing
J.~Niset, J.~Fiur\'a\ifmmode~\check{s}\else \v{s}\fi{}ek, and N.~J. Cerf,
  ``No-go theorem for gaussian quantum error correction,'' \emph{Phys. Rev.
  Lett.}, vol. 102, p. 120501, Mar 2009. [Online]. Available:
  \url{https://link.aps.org/doi/10.1103/PhysRevLett.102.120501}
\BIBentrySTDinterwordspacing

\bibitem{PJCC13}
\BIBentryALTinterwordspacing
S.~Pandey, Z.~Jiang, J.~Combes, and C.~M. Caves, ``Quantum limits on
  probabilistic amplifiers,'' \emph{Phys. Rev. A}, vol.~88, p. 033852, Sep
  2013. [Online]. Available:
  \url{https://link.aps.org/doi/10.1103/PhysRevA.88.033852}
\BIBentrySTDinterwordspacing

\bibitem{peggQuantumScissor}
D.~T. Pegg, L.~S. Phillips, and S.~M. Barnett, ``Optical state truncation by
  projection synthesis,'' \emph{Physical review letters}, vol.~81, no.~8, p.
  1604, 1998.

\bibitem{covarianceMatrixAlgebra}
G.~Spedalieri, C.~Ottaviani, and S.~Pirandola, ``Covariance matrices under
  bell-like detections,'' \emph{Open Systems \& Information Dynamics}, vol.~20,
  no.~02, p. 1350011, 2013.

\bibitem{WGHR21}
\BIBentryALTinterwordspacing
M.~S. Winnel, J.~J. Guanzon, N.~Hosseinidehaj, and T.~C. Ralph, ``Overcoming
  the repeaterless bound in continuous-variable quantum communication without
  quantum memories,'' 2021. [Online]. Available:
  \url{https://arxiv.org/abs/2105.03586}
\BIBentrySTDinterwordspacing

\end{thebibliography}
\bibliographystyle{IEEEtran}



\end{document}